\documentclass [twocolumn,floatfix,showpacs,amsmath,letterpaper] {revtex4}
\usepackage{amssymb}
\usepackage{amsmath}
\usepackage{amsfonts}
\usepackage{graphicx}
\usepackage{dcolumn}
\usepackage{bm}
\usepackage{color}
\usepackage{verbatim}
\begin{document}


\title{The longitudinal polarization of hyperons in the forward region in polarized  $pp$ collisions}
\author{Wei Zhou, Shan-shan Zhou and Qing-hua Xu}
\address{School of Physics, Shandong University, Jinan, Shandong 250100, China}


\vspace*{0.3cm}

\begin{abstract}
We study the longitudinal polarization of hyperons and anti-hyperons at forward
pseudorapidity, $2.5<\eta<4$, in singly polarized $pp$
collisions at RHIC energies by using different
parameterizations of the polarized parton densities and different
models for the polarized fragmentation functions. The results show
that the $\Sigma^+$ polarization is able to distinguish
different pictures on spin transfer in high energy fragmentation
processes; and the polarization of $\Lambda$ and
$\bar\Lambda$ hyperons can provide sensitivity to the helicity distribution
of strange sea quarks. The influence from beam remnant to
hyperon polarization in the forward region is also discussed.

\end{abstract}

\maketitle

The polarizations of hyperons and anti-hyperons in polarized proton-proton
 ($pp$) collisions
have attracted much attention recently, as they provide accesses to 
different aspects of spin physics~\cite{RHIC,Abelev:2009xg,deFlorian:1998ba,Boros:2000ex,Ma:2001na,Xu:2002hz,Xu:2005ru}. 
The running of high energy polarized $pp$ collider at RHIC
 provides a good opportunity for these studies~\cite{RHIC}.
The available experimental and theoretical studies mostly focused on
mid-rapidity region with high transverse momentum $p_T$~\cite{Abelev:2009xg,deFlorian:1998ba,Boros:2000ex,Ma:2001na,Xu:2002hz,Xu:2005ru,Chen:2007tm}. 
Phenomenological study in this region showed that $\Lambda$,$\Sigma^\pm$ 
hyperon production at high $p_T$ are dominated
by quark fragmentation, and their polarizations can be used to
study the polarized fragmentation functions~\cite{deFlorian:1998ba,Boros:2000ex,Ma:2001na,Xu:2002hz}.
On the other hand, the polarizations of $\bar\Lambda$ and other anti-hyperons, can provide sensitivity to polarized parton distribution functions,
in particular to those for the sea quark and anti-quark~\cite{Xu:2005ru,Chen:2007tm}.
Similar conclusion was also obtained for hyperons polarization in
semi-inclusive deep inelastic scattering~\cite{Zhou:2009mx,Kotzinian:2009gi}.
First measurement on longitudinal polarization of $\Lambda$ and
$\bar\Lambda$ in $pp$ collisions has been made in mid-rapidity 
at $\sqrt s=200$ GeV by the STAR experiment at RHIC at intermediate $p_T$
recently~\cite{Abelev:2009xg}. 
The experimental uncertainties are still statistics limited, 
and may be improved in future measurement. 
It is interesting that the STAR experiment is proposing a detector
upgrade in the near future, which may enable
hyperon polarization measurement in the forward range of 
$pp$ collisions~\cite{Hank:2009ags}.
Compared with hyperons at mid-rapidity, the forward hyperons carry
larger momentum fraction of the beam and their polarizations may have different
properties. 
Therefore, it is of practical importance to see the
behaviors of hyperon polarization in this kinematic region and this
short note aims to make such an analysis. 
We extend our calculations on hyperon polarization at mid-rapidity 
in $pp$ collisions to large rapidity region, and the results may 
serve as a guide for future experimental study.


We consider the inclusive production of hyperons
$H$ in $pp$ collisions with one beam longitudinally
polarized, where the polarization of hyperons is defined as,
\begin{equation}
P_H\equiv \frac{d\sigma_{p_+p\to H_+X}-d\sigma_{p_+p\to H_-X}}{d\sigma_{p_+p\to H_+X}+d\sigma_{p_+p\to H_-X}}
\equiv \frac{d\Delta\sigma}{d\sigma},
\end{equation}
where the subscripts +
and $-$ denote positive and negative helicity, and $d\Delta\sigma$ and
$d\sigma$ are the polarized and unpolarized inclusive production
cross sections.
Recently, the STAR experiment has measured the $\Lambda$
($\bar\Lambda$) production cross section in $pp$ collision with $p_T$
 up to 5 GeV at $\sqrt s=200$ GeV, which is in agreement
with pQCD calculation under factorization framework~\cite{Abelev:2006cs}. 
Similarly, the polarized cross section in Eq.(1) can be
 described by
 convolution of polarized parton distribution and fragmentation functions,
and the polarized cross section of partonic scattering which is calculable
in pQCD.
The hyperon polarization in $pp$ collision thus provide
connection to the polarized parton distribution and fragmentation
functions. 
The polarized parton distribution function can be in general obtained by
global parameterization of polarized data in different reactions, though
large uncertainties still exist especially for gluon and sea quarks
due to the lack of abundant data~\cite{deFlorian:2008mr}.

On the polarized fragmentation function, there are mainly two
approaches in literature. 
Similar as the (polarized) parton distribution function, one can 
make a global parameterization~\cite{deFlorian:1998ba}. 
However, the available data are still far from being abundant enough
to give solid constraint. 
At this stage, some key features still rely on phenomenological models 
\cite{Gustafson:1992iq,Boros:1998kc,Kot98,MSY,Anselmino:2000ga,Ellis2002}. 
In this analysis, we model the polarized fragmentation function according
to the origins of hyperon. 
The calculation method of hyperon polarization with this model 
has been applied to
the studies of hyperons and anti-hyperons polarization in
$e^+e^-$~\cite{Boros:1998kc,Liu:2000fi}, in polarized lepton-nucleon
scattering~\cite{Zhou:2009mx,Liu:2001yt,Zuotang:2002ub} and also in polarized $pp$
collisions~\cite{Xu:2002hz,Xu:2005ru,Chen:2007tm}. 
The detailed description for the case of $pp$ collisions can be found in
Ref.~\cite{Chen:2007tm}. In the following part of this paper, we
first briefly summarize the main points on the calculation procedure
of hyperon polarization in $pp$ collisions and then give the
results.

In our analysis, the Lund string fragmentation model~\cite{pythia} is used
 for the fragmentation process, which enables us to distinguish between $H$
that contain the fragmenting parton and those that do not. 
Those directly produced hyperons that contain the fragmenting parton can
be polarized, and the relation between hyperon polarization and the
polarization of fragmenting parton is described with two pictures of
hyperon spin content, which are obtained from the SU(6) wave function or
polarized deep-inelastic lepton-nucleon scattering data(DIS). Those
hyperons that do not contain the fragmenting parton are not
polarized. This model can also distinguish between directly produced
$H$ and $H$ which are decay products of heavier
resonances. 
The spin transfer in the decay process 
is also considered~\cite{Gustafson:1992iq,Boros:1998kc}.
One advantage of this model is that the main
feature of hyperon polarization is determined by the relative
contribution of different classes based on the origins, which is independent of
polarization property. 

In our calculation of hyperon polarization, we have used the {\sc
pythia} event generator for the unpolarized hyperon production in
$pp$ collision (see detailed description in Sec.IIC of Ref.~\cite{Chen:2007tm}). {\sc pythia} incorporates the hard scattering
processes and uses the Lund string fragmentation
model~\cite{pythia}, which has been commonly used for hadron-hadron
collisions and its output has been tested and tuned to describe a
vast body of data.


We first look into the different fractional contributions from the fragmentation of different quark flavors and gluon, which are expected to
play different roles in hyperon's polarization.
These contributions are independent of polarization property and have been determined by large amount of data collected over the past years.
They are thus considered to be well-modeled in Monte-Carlo event generators like {\sc pythia}.
We have used {\sc pythia}6.420~\cite{pythia} in its default tune to estimate these contributions.
We note that the recent data on $\Lambda$($\bar\Lambda$) production at RHIC in
$pp$ collisions at $\sqrt s=$200 GeV can also be described
by {\sc pythia} with certain $K$ factor~\cite{Abelev:2006cs}, 
which does not affect the relative fractions here and thus 
the corresponding hyperon polarization shown later.

\begin{figure}[!hptb]
\includegraphics[width=0.48\textwidth]{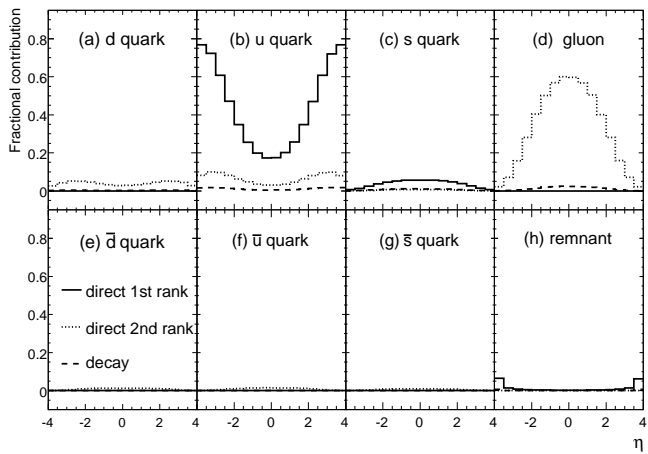}
\caption{Contributions to $\Sigma^+(uus)$ production with
$p_T\ge 3$ GeV in $pp$ collisions at $\sqrt{s}=500$ GeV versus $\eta$.}
\label{fig:sigmap500}
\end{figure}

\begin{figure}[!hptb]
\includegraphics[width=0.48\textwidth]{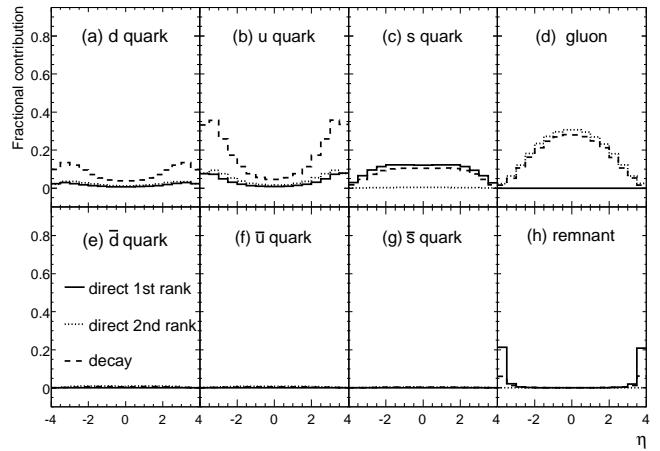}
\caption{Contributions to $\Lambda(uds)$ production with
$p_T\ge 3$ GeV in $pp$ collisions at $\sqrt{s}=500$ GeV versus $\eta$.}
\label{fig:lambda500}
\end{figure}

Figure~\ref{fig:sigmap500} shows the results of the fractional
contributions to $\Sigma^+$ production from the fragmentation of
antiquarks/quarks of different flavors and of gluons in $pp$
collisions versus pseudorapidity $\eta$ at $\sqrt{s}=500$ GeV 
for hyperon transverse momenta $p_T>3$ GeV.
We see that, $\Sigma^+$ production at large $\eta$ is dominated
by $u$ quark fragmentation.
Most of them are directly produced and
contain the fragmenting $u$ quark (denoted as ``direct 1st rank'' in Fig.1),
whose percentage reaches 80\% at $|\eta|=4$.
The contribution from gluon is very small at $|\eta|>2.5$.
Most of $\Sigma^+$'s from gluon 
 are directly produced and apparently don't contain
the fragmenting gluon (denoted as ``direct 2nd rank'' in Fig.1).
The decay contribution from heavier hyperons (denoted as ``decay'' in Fig.1)
is very small in the whole $\eta$ region.
The contributions from sea quarks are negligible as expected.
In the large $\eta$ region, we can also see
the contribution of beam remnant, as seen in Fig. 1(h).
As $u$ quark carries most of $\Sigma^+$'s spin either in SU(6) or DIS picture,
we expect $\Sigma^+$'s polarization be sizable and 
increase with $\eta$.

The fractional contributions for $\Lambda$ production with
$p_T$$>$3 GeV at $\sqrt{s}=500$ GeV are shown in
Fig.~\ref{fig:lambda500}. 
Comparing with the results of $\Sigma^+$, 
the situation for $\Lambda$ is more complicated. 
The largest contribution comes from $u$ quark's fragmentation in the
forward region of 2.5$<$$\eta$$<$4, but most of them are from decay
contribution. 
The shape of $d$ quark's contribution is similar as that of $u$ quark, 
but the size is about one half. 
The contribution from decays of heavier hyperons is
sizable for each flavor in the whole $\eta$ region. The contribution
of $s$ quark is a few percent in the forward region, and thus
$\Lambda$ polarization is expected to be small as $u$ or $d$ quark
does not carry $\Lambda$'s spin in SU(6) picture or carries only 
a small fraction in DIS picture. 
At $\eta\approx 4$, the contribution of beam remnant
is also seen as $\Sigma^+$'s case.

\begin{figure}[!hptb]
\includegraphics[width=0.46\textwidth]{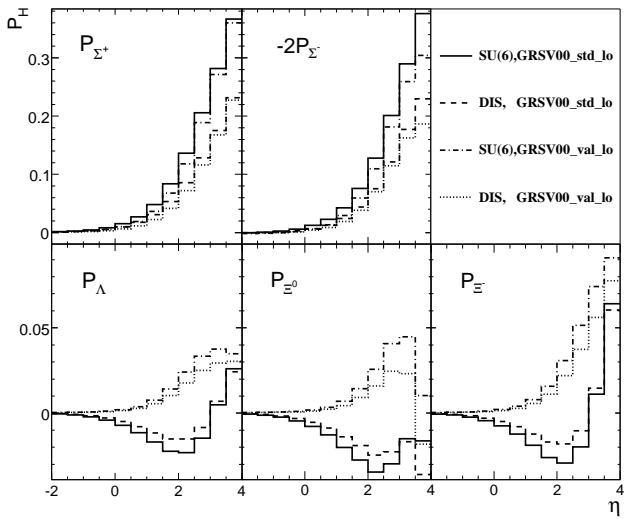}
\caption{Longitudinal polarization for $\Lambda$, $\Sigma^+$, $\Sigma^-$, $\Xi^0$
and $\Xi^-$ hyperons
with $p_T\ge 3$ GeV in $pp$ collisions at $\sqrt s=500$ GeV
with one beam longitudinally polarized versus $\eta$.
Positive $\eta$ is taken along the direction of the polarized beam.
}
\label{fig:polh500eta}
\end{figure}

Figure~\ref{fig:polh500eta} shows the evaluated polarization results for the $\Lambda$, $\Sigma^+$, $\Sigma^-$, $\Xi^0$
and $\Xi^-$ hyperons as a function of pseudorapidity $\eta$ for $p_T>$3 GeV at $\sqrt{s}=500$ GeV using different parameterizations for the polarized parton distributions and using the SU(6) and DIS pictures for the spin transfer factors in the fragmentation.
To be consistent with our model on polarized fragmentation process, the polarized
partonic cross section is evaluated at leading order (LO)~\cite{bms,Xu:2002hz} and LO sets of polarized parton distribution functions are used. 
Here we pick two LO sets of GRSV00 parameterization for the polarized parton distribution functions, to study the sensitivity of hyperon polarization to strange sea quark helicity distributions,
as the main difference between these two sets is on the sea quarks~\cite{Gluck:2000dy}.
The main characteristics of the polarization results in the forward region of 2.5$<\eta<$4 are:

\begin{itemize}
\item The magnitudes of $\Sigma^+$, $\Sigma^-$ polarization are much larger 
      than those for
      the $\Lambda$, $\Xi^0$ and $\Xi^-$ hyperons because of the dominant
      contributions of the valence quarks ($u$ and $d$) in their
      productions. The $\Sigma^+$ polarization increases with
      $\eta$ and is larger than 0.3 at $\eta\approx 4.0$ with the SU(6) picture.      We take into account the contribution of beam remnant in the
      large $\eta$ region, using a simple model based on
      SU(6) wave-function~\cite{Zuotang:2002ub}.
\item
      The polarizations of $\Sigma^+$ and $\Sigma^-$ are
      sensitive to different pictures of spin transfer in
      fragmentation process, and thus can distinguish whether SU(6) or DIS
      picture is suitable for the spin transfer in fragmentation.
      The results for $\Sigma^+$ and $\Sigma^-$ differ in sign
      because of the sign difference between $\Delta u(x)$ and $\Delta d(x)$,
      and the magnitude of $\Sigma^-$ polarization is about one half of 
      $\Sigma^+$ polarization due to the size difference between $\Delta u(x)$ 
      and $\Delta d(x)$.
\item  The results for $\Lambda$, $\Xi^0$ and $\Xi^-$ hyperons are in general smaller compared with $\Sigma^\pm$ polarization as their contributions of $s$ quark are smaller than that of $u,d$ quarks to $\Sigma^\pm$.
 Their polarizations have similar shapes at $\eta<3$ because 
 they are all dominated by the $s$ quark's contribution,
 and the different shapes at $\eta\approx 4$ come from the different
 effects of beam remnants. 
 \item The polarizations of $\Lambda$, $\Xi^0$ and $\Xi^-$ are sensitive to different choices of polarized parton distribution functions rather than different spin transfer pictures in fragmentation.
The sensitivity mainly comes from the strange sea quark's contribution.
Their polarizations can thus provide sensitivity to the 
helicity distribution function of strange quark $\Delta s(x)$ with 
hyperon $\eta$ up to 3.5.
At even larger $\eta$, the sensitivity
decreases as the contribution of $s$ quark drops to zero
and the beam remnant's contribution shows up.

\end{itemize}

\begin{figure}[!hptb]
\includegraphics[width=0.46\textwidth]{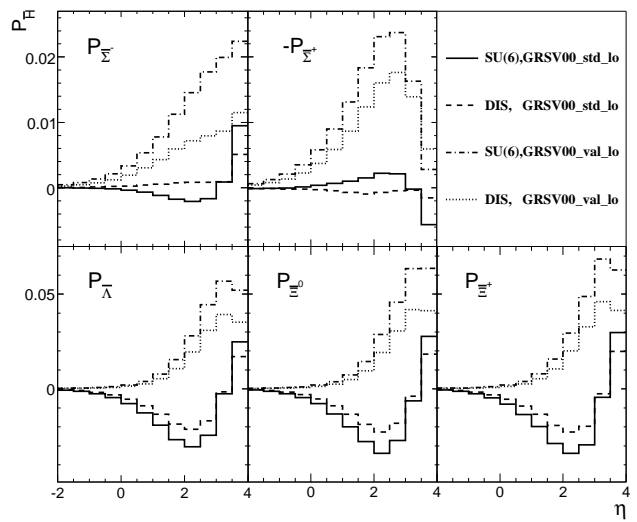}
\caption{
The same as Fig.3 for the polarizations of anti-hyperons $\bar \Sigma^-$, $\bar \Sigma^+$, $\bar \Lambda$, $\bar \Xi^0$ and $\bar \Xi^+$. 
}
\label{fig:polah500eta}
\end{figure}

With the same procedure, we also evaluated the polarizations for
anti-hyperons in the forward region and
the results are shown in Fig.~\ref{fig:polah500eta}.
The general shapes of their polarizations are quite similar as those
obtained with higher $p_T>8$ GeV in mid-rapidity~\cite{Chen:2007tm}, 
but the magnitudes are smaller. 
The polarizations for $\bar{\Lambda}$, $\bar{\Xi}^0$ and 
$\bar{\Xi}^+$ at $2.5<\eta<4$ are similar both in size and in shape. 
Unlike their anti-particles, the contribution from beam remnant is negligible here.
Their polarizations are all sensitive to different sets of polarized parton distribution functions as their polarizations are dominated by contribution of $\bar s$ quark.
The polarizations of $\bar{\Lambda}$, $\bar{\Xi}^0$ and $\bar{\Xi}^+$ thus
provide sensitivity to $\Delta {\bar s}(x)$.
The size of $\bar \Sigma^-$ and $\bar \Sigma^+$ polarization is smaller than that for the $\bar\Lambda$ and $\bar\Xi$ hyperons, and their
polarizations are dominated by the contributions from $\bar{u}$ and $\bar{d}$ quarks.
The $\bar \Sigma^-$ and $\bar \Sigma^+$ polarizations show sensitivity to $\Delta \bar u(x)$ and $\Delta\bar d(x)$ with $\eta$ up to 4.
The results for $\bar \Sigma^-$ and $\bar \Sigma^+$ polarizations differ in sign 
because of the sign difference in $\Delta \bar u(x)$ and $\Delta\bar d(x)$ in the
 ``valence'' set of GRSV2000.

\begin{figure}[!hptb]
\includegraphics[width=0.48\textwidth]{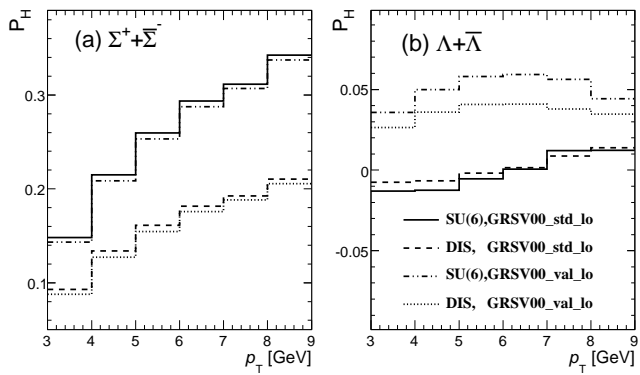}
\caption{Longitudinal polarization for $\Sigma^+ + \bar{\Sigma}^-$ and  $\Lambda + \bar{\Lambda}$ versus $p_T$ with $2.5<\eta<3.5$ in $pp$ collisions at $\sqrt s=500$ GeV with one beam longitudinally polarized. 
}
\label{fig:pols500pt}
\end{figure}

We also make calculations for (anti-)hyperons with $p_T$$>$2 GeV 
at $\sqrt{s}=200$ GeV.
The main features of their polarizations remain the same as the above results at
 $\sqrt s$=500 GeV, and just
the size of the polarizations is slightly smaller.
 This is because of the smaller fractional
contributions from the quarks. In addition, in this case the
contribution from beam remnant becomes larger at $\eta>3$ as the
$p_T$ is a bit lower, and this may introduce some uncertainty to the
evaluations of their polarizations.

In the case that the separation of hyperon and anti-hyperon is
difficult with forward detectors as proposed in the STAR detector 
upgrade~\cite{Hank:2009ags}, we also make corresponding
calculations by combining hyperon and its anti-particle. 
The combined
results of polarization are shown in Fig.~\ref{fig:pols500pt} for
$\Sigma^+ + \bar{\Sigma}^-$ and $\Lambda + \bar{\Lambda}$
versus $p_T$ for 2.5$<\eta<$3.5  
in $pp$ collisions at $\sqrt s=500$ GeV.
It shows that a precision measurement of
$\Sigma^+ + \bar{\Sigma}^-$ polarization is able to
distinguish different models for the spin transfer in fragmentation
process, and the polarization of $\Lambda + \bar{\Lambda}$ may
provide useful information for the helicity distribution functions
of strange sea quarks.


In summary, we have investigated the longitudinal polarizations of
hyperons and anti-hyperons in polarized $pp$ collisions in the
forward region of 2.5$<\eta<$4. The results show that $\Sigma^+$
polarization in this kinematic region is significant and can
distinguish the two spin transfer pictures SU(6) and DIS in 
the fragmentation process. The polarizations of $\Lambda$ and
$\bar\Lambda$ can provide sensitivity to strange sea and anti-sea
quark polarization in polarized nucleon. Precision measurements at
the RHIC $pp$ collider should be able to test these predictions and
thus provide new insights into the polarized fragmentation process and
the strange sea quark polarization in the nucleon.

We thank Prof. Zuo-tang Liang for the helpful discussions.
This work was supported in part by the National Natural Science
Foundation of China under the approval Nos. 10525523 \& 10975092, and
by SRF for ROCS, SEM.



\begin{thebibliography}{99}

\bibitem{RHIC}
  G.~Bunce, N.~Saito, J.~Soffer and W.~Vogelsang,
  Ann.\ Rev.\ Nucl.\ Part.\ Sci.\  {\bf 50}, 525 (2000).

\bibitem{Abelev:2009xg}
  B.~I.~Abelev {\it et al.}  [STAR Collaboration],
  Phys.\ Rev.\  D {\bf 80}, 111102 (2009).

\bibitem{deFlorian:1998ba}
  D.~de Florian, M.~Stratmann and W.~Vogelsang,
  Phys.\ Rev.\ Lett.\  {\bf 81}, 530 (1998); Phys. Rev. D{\bf 57}, 5811 (1998).

\bibitem{Boros:2000ex}
 C.~Boros, J.T.~Londergan and A.W.~Thomas,
 Phys.\ Rev.\  D {\bf 62}, 014021 (2000).

\bibitem{Ma:2001na}
  B.Q.~Ma, I.~Schmidt, J.~Soffer and J.J.~Yang,
  Nucl.\ Phys.\  A {\bf 703}, 346 (2002).

\bibitem{Xu:2002hz}
  Q.~H.~Xu, C.~X.~Liu and Z.~T.~Liang,
  Phys.\ Rev.\  D {\bf 65}, 114008 (2002);
%
  Q.~H.~Xu and Z.~T.~Liang,
  {\it ibid.} {\bf 70}, 034015 (2004).

\bibitem{Xu:2005ru}
  Q.~H.~Xu, Z.~T.~Liang and E.~Sichtermann,
  Phys.\ Rev.\  D {\bf 73}, 077503 (2006).

\bibitem{Chen:2007tm}
  Y.~Chen, Z.~T.~Liang, E.~Sichtermann, Q.~H.~Xu and S.~S.~Zhou,
  Phys.\ Rev.\  D {\bf 78}, 054007 (2008).

\bibitem{Zhou:2009mx}
  S.~S.~Zhou, Y.~Chen, Z.~T.~Liang and Q.~H.~Xu,
  Phys.\ Rev.\  D {\bf 79}, 094018 (2009).

\bibitem{Kotzinian:2009gi}
  A.~Kotzinian, 
  arXiv:0907.3270 [hep-ph].

\bibitem{Hank:2009ags}
   H.~Crawford,
   ``STAR Future Plans and Upgrades'',
   Presentation at 2009 RHIC \& AGS Annual Users' Meeting; 
  L.~Nogach, talk at DSPIN-09, 
  arXiv:0911.4388 [hep-ex].

\bibitem{Abelev:2006cs}
  B.~I.~Abelev {\it et al.}  [STAR Collaboration],
  Phys.\ Rev.\  C {\bf 75}, 064901 (2007).

\bibitem{deFlorian:2008mr}
  D.~de Florian, R.~Sassot, M.~Stratmann and W.~Vogelsang,
  Phys.\ Rev.\ Lett.\  {\bf 101}, 072001 (2008).

\bibitem{Gustafson:1992iq}
  G.~Gustafson and J.~Hakkinen,
  Phys.\ Lett.\  B {\bf 303}, 350 (1993).

\bibitem{Boros:1998kc}
  C.~Boros and Z.~T.~Liang,
  Phys.\ Rev.\  D {\bf 57}, 4491 (1998).

\bibitem{Kot98} 
    A. Kotzinian, A. Bravar and D. von Harrach, 
    Eur. Phys. J. C{\bf 2}, 329 (1998).

\bibitem{MSY}
   B.~Q.~Ma, I.~Schmidt and J.~J.~Yang, Phys.\ Rev.\ D 
   {\bf 61}, 034017 (2000); {\bf 63}, 037501 (2001);
   B.~Q.~Ma, I.~Schmidt, J.~Soffer and J.~J.~Yang, 
   Phys.\ Rev.\ D {\bf 62}, 114009 (2000).

\bibitem{Anselmino:2000ga}M.~Anselmino, M.~Boglione and F.~Murgia,
  Phys.\ Lett.\  B {\bf 481}, 253 (2000).

\bibitem{Ellis2002} J.~R.~Ellis, A.~Kotzinian and D.~V.~Naumov,
     Eur.\ Phys.\ J.\ C {\bf 25}, 603 (2002); 
     J.~Ellis, A.~Kotzinian, D.~Naumov and M.~Sapozhnikov, 
     Eur. Phys. J. C {\bf 52},283 (2007).

\bibitem{Liu:2000fi}
  C.~X.~Liu and Z.~T.~Liang,
  Phys.\ Rev.\  D {\bf 62}, 094001 (2000).

\bibitem{Liu:2001yt}
  C.~X.~Liu, Q.~H.~Xu and Z.~T.~Liang,
  Phys.\ Rev.\  D {\bf 64}, 073004 (2001);
%
  H.~Dong, J.~Zhou and Z.~T.~Liang,
  {\it ibid.} {\bf 72}, 033006 (2005).

\bibitem{Zuotang:2002ub}
  Z.~T.~Liang and C.~X.~Liu,
  Phys.\ Rev.\  D {\bf 66}, 057302 (2002).
 

\bibitem{pythia} T. Sj\"{o}strand, S. Mrenna, P. Skands, JHEP {\bf 0605}, 026 (2006); B. Andersson, G. Gustafson, G. Ingelman and T. Sj\"{o}strand, Phys. Rep. {\bf 97}, 31 (1983).
 
\bibitem{bms}J. Babcock, E. Monsay, D. W. Sivers, Phys. Rev. Lett. {\bf 40}, 1161 (1978); \ Phys. Rev. D {\bf19}, 1483 (1979).

\bibitem{Gluck:2000dy}
  M.~Gluck, E.~Reya, M.~Stratmann and W.~Vogelsang,
  Phys.\ Rev.\  D {\bf 63}, 094005 (2001).



\end{thebibliography}
\end{document}